# Topological Origins of Flexibility and Internal Stress in Sodium Aluminosilicate Glasses

Ernest Ching[1], Mathieu Bauchy[1*]

[1]Physics of AmoRphous and Inorganic Solids Laboratory (PARISlab), University of California, Los Angeles, CA 90095-1593, U.S.A.
*Corresponding author: Prof. Mathieu Bauchy, bauchy@ucla.edu

**ABSTRACT**

In the framework of topological constraint theory, network glasses are classified as flexible, stressed–rigid, or isostatic if the number of atomic constraints is smaller, larger, or equal to the number of atomic degrees of freedom. Here, based on molecular dynamics simulations, we show that sodium aluminosilicate glasses exhibit a flexible-to-stressed–rigid transition driven by their composition. This transition manifests itself by a loss of atomic mobility and an onset of internal atomic stress. Importantly, we find that the flexible-to-rigid (i.e., loss of internal flexibility) and unstressed-to-stressed transitions (i.e., onset of internal stress) do not occur at the same composition. This suggests that the isostatic state (i.e., rigid but unstressed) is achieved within a window rather than at a threshold composition.

## I. INTRODUCTION

Sodium aluminosilicate glasses are well recognized for their outstanding mechanical properties and its ability to undergo ion-exchange [1–4] with larger alkali elements, such as potassium, which increases its surface compressive strength. As such, sodium aluminosilicate glasses find a wide range of commercial applications, including its use as smartphone display surfaces, e.g. Corning® Gorilla® Glass [5,6] and as substrates for organic electronics [7]. In order to better understand the post-ion exchange properties of aluminosilicate glasses, this paper examines the origins of flexibility and internal stress in sodium aluminosilicate glasses within the framework of topological constraint theory (TCT) [8–15]. TCT describes the rigidity of a glass network by modeling atoms as truss nodes, and chemical bonds as truss members [10]. The number of constraints per atom ($n_c$) is the sum of the following two types of constraints: radial bond stretching constraints and angular bond bending constraints, which varies between different atomic types based on the geometry of local connectivity. TCT allows for three modes: (i) flexible ($n_c < 3$), where the number of inter-atomic constraints is fewer than the number of atomic degrees of freedom, (ii) isostatic ($n_c = 3$), where the number of constraints exactly balance the degrees of freedom, and (iii) stressed-rigid ($n_c > 3$), where the network is over-constrained, leading to the development of internal stresses [16].

The atomic structure and physical properties, including density, stiffness, thermal expansion, and corrosion behavior of sodium aluminosilicate glasses have been well studied in previous experiments and simulations [17–21]. However, it should be noted that per-aluminous glasses with high alumina content have not been studied extensively in experiments due to decreased glass formation ability with increased alumina content [22]. Hence, this study relies on molecular dynamics simulations in order to simulate alumina-rich glasses—this study examines compositions of $(Na_2O)_{30}(SiO_2)_{70-x}(Al_2O_3)_x$, with $x$ = 0% to 70% at 5% increments. Moreover, although topological constraint models have been proposed for different glasses, e.g., sodium silicate [23], sodium phosphosilicate [24], and soda lime borate [25], no comprehensive topological constraint model currently exist for sodium aluminosilicate glasses.

Hence, this paper aims to fill this gap of knowledge by proposing a topological constraint model (TCM) for sodium aluminosilicates, which is compared to and verified with signatures of flexibility and internal stress obtained from simulations.

## II. SIMULATION METHODOLOGY

### A. Preparation of the glasses

Sodium aluminosilicate glasses with compositions of $(Na_2O)_{30}(SiO_2)_{70-x}(Al_2O_3)_x$, with $x$ = 0% to 70% at 5% increments are simulated by molecular dynamics using the LAMMPS [26] package. All simulations are performed using the well-established empirical potential parametrized by Teter [27], which has been verified to predict realistic structural, mechanical and dynamical properties for sodium aluminosilicate glasses [18,28–32]. The short-range interactions are modeled by a Buckingham potential with a cutoff of 8.0 Å, and the Coulombic interactions are evaluated using the Ewald method, with a cutoff of 12.0 Å [28].

All glass compositions are composed of roughly 3000 atoms and the initial configuration is created by randomly placing atoms in a cubic simulation box while avoiding any unrealistic overlap. The cooled glasses are then formed by (i) creating the melts at 4000 K with a Gaussian distribution to lose the memory of the initial configuration, (ii) equilibrating the melts at 4000 K and 2 GPa for 100 ps, and then at 4000 K and zero pressure for an additional 100 ps to lose the memory of initial configurations, (iii) cooling from 4000 K to 300 K at 1 K/ps at zero pressure, and (iv) relaxing the cooled glass at 300 K and zero pressure for 200 ps. The glass forming procedure described above is performed in the *NPT* ensemble with a Nosé–Hoover thermostat and barostat [33,34]. A timestep of 1 fs is used for all simulations. All glass compositions are repeated for 6 trials by using different random velocities for the creation of glass melts.

### B. Structure characterization

Determining the radial and angular constraint of different atomic sub-species is important in creating a realistic topological constraint model. To this end, each as-cooled glass composition is relaxed at 300 K for 0.1 ns in the *NVT* ensemble. An atomic trajectory data file is obtained during this relaxation, which is used to analyze the structure of the glass using MATLAB [35].

Each oxygen atom is classified as a non-bridging oxygen (NBO), bridging oxygen (BO) or a tricluster oxygen (TO) based on the number of network formers (silicon and aluminum) each oxygen atoms is connected to. Similarly, each aluminum atom is classified as a four-fold aluminum ($Al^{IV}$) or a five-gold aluminum ($Al^V$) based on the number of oxygen atoms each aluminum is connected to. A MATLAB script is used to perform the above classification, which, for each oxygen atom, calculates the distance between the oxygen atom and each silicon or aluminum atom, and adds up the number of silicon or aluminum atoms that are less than a specified cutoff distance from the oxygen. The O–Si/Al cutoff is determined by examining the O–Si and O–Al partial radial distribution functions (PDF) in OVITO [36], and choosing the first minimum in the PDF, to a precision of 0.05 Å, as the cutoff [37,38]. It is found that the cutoffs vary between 2.0 Å (for $x = 0\%$) and 2.2 Å (for $x = 70\%$) depending on the composition of the glass since O–Si/Al bond lengths increase as the percentage of alumina increases. Having classified each O and Al atom into their respective sub-species, the bond angle distribution for each sub-species of O and Al is obtained by calculating the angles between each pair of bonds within the cutoff of each central O or Al atom.

**C. Mean square displacement computation**

The mean square displacement of a glass after an energy bump can be used as a measure of flexibility. In this investigation, the as-cooled glasses were first cooled to from 1 K to $10^{-7}$ K in 20 ps in the *NVT* ensemble. An energy minimization is performed using the Polak-Ribiere version of the conjugate gradient (CG) algorithm [39] with energy and force tolerances of $10^{-8}$ and $10^{-8}$ kcal/mol·Å, respectively. Then, the system is subjected to a 200 meV energy bump—applied in the form of an increment in the kinetic energy of the atoms, following a Gaussian velocity distribution. Based on the equipartition theorem, half of this energy bump eventually results in an increase in the potential energy of the atoms, while the other half yields an increase in temperature. This energy bump corresponds to a final system temperature of 750 K, which is below the glass transition temperature of the simulated glasses [17,19] to ensure that the glasses remain in the solid phase. The glasses are then allowed to relax in the *NVE* ensemble for 2 ns, and the mean square displacement of all atoms is computed at every 1 ps. 6 trials are obtained for each composition by using different random velocities for the energy bump.

**D. Internal stress computation**

Stress is normally defined for macroscopic objects or ensembles of atoms as the energy per unit volume, but ill-defined for individual atoms due to the lack of a clear physical interpretation. Nevertheless, to quantify the magnitude of the local forces acting on individual atoms, the "stress per atom" framework formulated by Thompson *et al.* [40] is adopted in our investigation. The stress per atom ($\sigma_i$) is defined in the following equation [41]:

$$3\sigma_i V_i = m_i v_i^2 + \vec{r_i} \cdot \vec{F_i} \qquad \text{(Eq. 1)}$$

where $i$ denotes the atom of interest, and $V_i$, $m_i$, $v_i$, and $\vec{r_i}$ denote the Voronoi volume, mass, velocity, and position of the atom, and $\vec{F_i}$ is the sum of all forces applied on the atom by all the other atoms in the system. Several previous studies have utilized this approach to quantify local stresses in an atomic network [42–46]. The computed stresses are either positive (in tension) or negative (in compression), and the sum of all stresses per atom multiplied by their respective Voronoi volumes equal zero since there is no external pressure applied to the glass macroscopically. The stress per atom is calculated in LAMMPS after bringing the glass to $10^{-7}$ K and performing the energy minimization procedure as described in Section II.C.

The internal stress per silicon atom ($\sigma_{\text{Si,internal}}$) is defined in Eq. 2 as the difference between the stress per silicon atom in the as-cooled glass ($\sigma_{\text{Si,glass}}$) and the stress per silicon atom in an isolated $Q^n$ cluster (an SiO$_4$ tetrahedron unit connected to $n$ bridging oxygens), henceforth termed "reference stress" ($\sigma_{\text{Si,reference}}$):

$$\sigma_{\text{Si,internal}} = \sigma_{\text{Si,glass}} - \sigma_{\text{Si,reference}} \quad \text{(Eq. 2)}$$

More details regarding this calculation can be found in Ref. [41].

### III. RESULTS AND DISCUSSION

#### A. Radial constraints

In order to construct a realistic topological constraint model, the number of radial constraints per atom is modeled by examining the connectivity of the glass. In this section, the oxygen atoms are classified into 3 sub-species, each of which creates different numbers of constraints: non-bridging oxygen (NBO), bridging oxygen (BO) and tricluster oxygen (TO) [47]. The aluminum atoms are classified into 2 sub-species: four-fold aluminum (Al$^{IV}$) and five-fold aluminum (Al$^{V}$) [48].

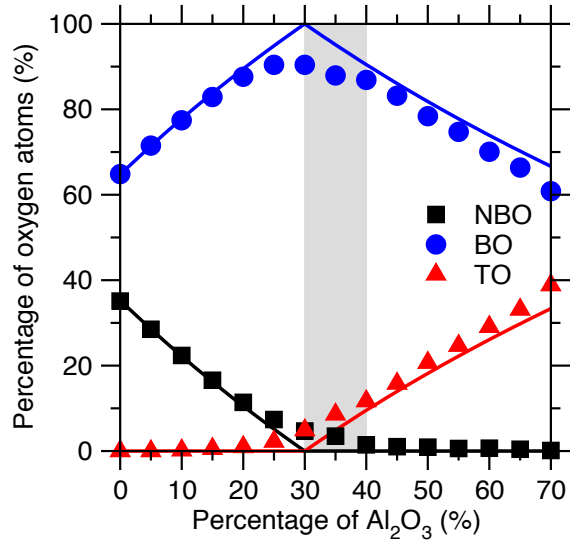

**FIG. 1:** Theoretical and simulated fractions of non-bridging oxygen (NBO), bridging oxygen (BO), and tricluster oxygen (TO) as a function of composition. The lines indicate the theoretical fractions as presented in TABLE 2, and the symbols indicate simulated fractions. The grey area indicates the region where the glass is isostatic.

For a pure sodium silicate ($x = 0\%$), each Na acts as a network modifier by breaking one Si–BO covalent bond and creating an Na–NBO ionic bond. Hence, the theoretical number of NBO is equal to the number of Na. As the percentage of alumina increases from 0% to 30%, each four-fold coordinated Al (Al$^{IV}$) removes one Na–NBO bond by taking an Na into its local vicinity in order to balance the excess unit negative charge on the AlO$_4$ tetrahedron. At $x = 30\%$, the number of Na exactly equal the number of Al, and the structure of the glass is comprised of SiO$_4$ and AlO$_4$ tetrahedrons, with all oxygen atoms acting as BO as seen in FIG. 1. At $x > 30\%$, as the ratio of Al/Na > 1, the newly added Al must rely on a different mechanism in order to maintain local charge neutrality. This is achieved by creating one TO (tricluster oxygen) for each non-sodium charge compensated Al, where each TO is connected to 3 network formers (Al or Si) [19]. The above model has been used in a number of studies [17,19,49], and the existence of TO has been verified in simulations and experiments [17,18,50]. Furthermore, simulated glasses in this study suggest that this model is fairly consistent with simulations; the model predicts the formation of NBO accurately for pre-aluminous glasses, but slightly under-predicts the fraction of TO for per-aluminous glasses.

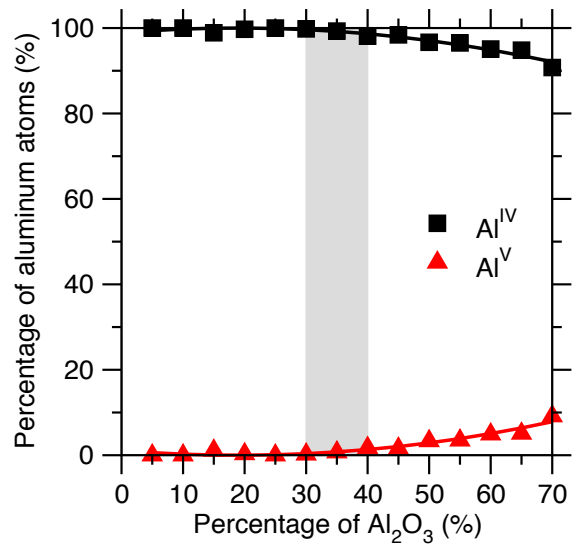

**FIG. 2:** Simulated fractions of four-fold aluminum (Al$^{IV}$) and five-fold aluminum (Al$^V$) as a function of composition. The lines serve as guides for the eyes. The grey area indicates the region where the glass is isostatic.

Since the fraction of Al$^V$ is hard to predict, this paper assumes a theoretical model where all Al are four-fold coordinated. In our simulations, the fraction of Al$^V$ is observed to increase as the percentage of alumina increases, up to a maximum of 10% at 70% alumina content as shown in FIG. 2. The existence of Al$^V$ has been observed in previous simulations and experiments [18,19,51], and are a consequence of increased network connectivity in alumina-rich glasses.

## B. Angular constraints

In addition to identifying the radial connectivity of each atomic sub-species, it is also important to assign the correct number of angular constraints to each sub-species. This can be achieved by examining the simulated bond angle distributions (BAD) of each sub-species. In general, distributions featuring sharp peaks suggest low angular bond mobility and wide distributions suggest high angular flexibility [52–55].

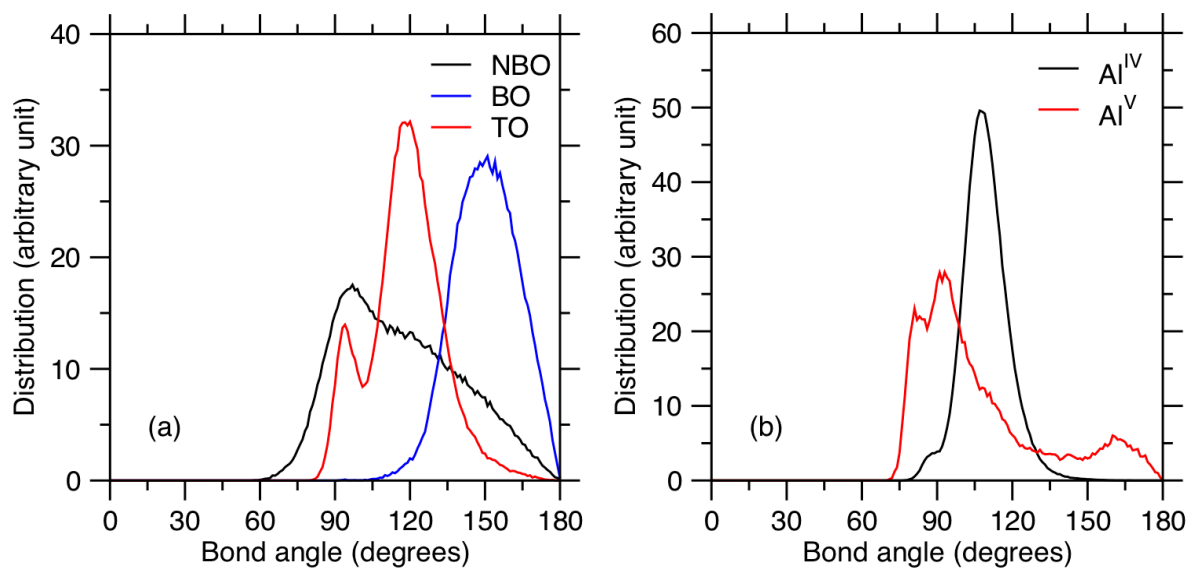

**FIG. 3: (a)** Simulated bond angle distribution for non-bridging oxygen (NBO) and bridging oxygen (BO) in $(Na_2O)_{30}(SiO_2)_{70}(Al_2O_3)_{00}$, and for tricluster oxygen (TO) in $(Na_2O)_{30}(SiO_2)_{00}(Al_2O_3)_{70}$. **(b)** Simulated bond angle distribution for four-fold aluminum ($Al^{IV}$) and five-fold aluminum ($Al^V$) in $(Na_2O)_{30}(SiO_2)_{00}(Al_2O_3)_{70}$.

Upon examination of FIG. 3(a), it is observed that the NBO BAD has a wide distribution between 60° and 180°, which suggests that the NBO creates no angular constraints. On the contrary, the BO BAD has a concentrated peak at 150°. The TO BAD has two concentrated peaks at 90° and 120°, which corresponds to edge-sharing and trigonal planar geometries respectively. These bond angles agree with previous simulations and experiments [18,56–60]. The narrow BADs of BO and TO suggest that BO and TO have low angular bond mobility.

For the BADs of different Al sub-species as shown in FIG. 3(b), $Al^{IV}$ shows a narrow peak at 109°, which corresponds to the expected tetrahedron geometry. However, $Al^V$ shows a wide distribution between 75° and 180° with peaks at 90° and 180°, which corresponds to a distorted octahedral geometry. These bond angles agree with previous simulations [18,19,51,56]. Due to the sharpness of their respective BADs, the $Al^{IV}$ is theorized to create angular constraints while the $Al^V$ is thought to create no angular constraints.

## C. Topological constraint model

**TABLE 1:** Assumptions used to calculate the theoretical atomic fractions for the construction of the proposed topological constraint model, where $x$ is the percentage of $Al_2O_3$, which varies from 0% to 70%. Percentage of atoms are expressed per unit of $(Na_2O)_{30}(SiO_2)_{70-x}(Al_2O_3)_x$.

| $x \leq 30\%$: | $x > 30\%$: |
|---|---|
| 1) No $Al^V$ are formed. | 1) No $Al^V$ are formed. |
| 2) No TO are formed. | 2) No NBO are formed. |
| 3) Each Al creates one charge compensating Na ($Na_{Al}$). | 3) Every Na is used to charge compensate an $Al^{IV}$, i.e. every Na is an $Na_{Al}$. |
| 4) Each remaining Na ($Na_{NBO}$) creates one ionic bond to an NBO. | 4) Each $Al^{IV}$, except those that are charge compensated by an $Na_{Al}$, create one TO. |

**TABLE 2:** Theoretical atomic fractions, bond stretching constraints, and bond bending constraints behind the proposed topological constraint model, where $x$ is the percentage of $Al_2O_3$, which varies from 0% to 70%. Percentage of atoms are expressed per unit of $(Na_2O)_{30}(SiO_2)_{70-x}(Al_2O_3)_x$.

| Atomic Type | Percentage ($x \leq 30\%$) | Percentage ($x > 30\%$) | Bond Stretching Constraints | Bond Bending Constraints | Total # of Constraints |
|---|---|---|---|---|---|
| NBO | $60 - 2x$ | 0 | $2/2$ | 0 | 1 |
| BO | $110 + 3x$ | $230 - x$ | $2/2$ | 1 | 2 |
| TO | 0 | $2x - 60$ | $3/2$ | 3 | $9/2$ |
| Si | $70 - x$ | $70 - x$ | $4/2$ | 5 | 7 |
| $Al^{IV}$ | $2x$ | $2x$ | $4/2$ | 5 | 7 |
| $Al^V$ | 0 | 0 | $5/2$ | 0 | $5/2$ |
| $Na_{NBO}$ | $60 - 2x$ | 0 | $1/2$ | 0 | $1/2$ |
| $Na_{Al}$ | $2x$ | 60 | 0 | 0 | 0 |

Typical radial bond stretching (BS) constraints, BS = $n/2$ [49], where $n$ denotes the connectivity of the central atom, are assumed for all atomic types except for $Na_{Al}$ since a charge compensating $Na_{Al}$ does not create a directional "bond" [61]. Moreover, $n$ for $Na_{NBO}$ is set to equal 1 instead of its coordination number (i.e., 6) [59]. Typical angular bond bending (BB) constraints, BB = $2n - 3$, are assumed for all atomic types except for NBO, $Al^V$, $Na_{NBO}$, and $Na_{Al}$. As explained in Section III.B, no bond bending constraints are assumed for NBO and $Na_{NBO}$ because a $Na_{NBO}$ ionically bonded to an NBO is not thought to impose any angular constraints. Again, a $Na_{Al}$ in the vicinity of an aluminum atom does not create a directional "bond" and therefore does not impose any angular constraints. As shown in **FIG. 3**, NBO and $Al^V$ have wide BADs, which suggests that no angular constraints are imposed by NBO or $Al^V$. The theoretical atomic fraction, BS, BB, and the total number of constraints per atom for each atomic sub-species is tabulated in **TABLE 2**.

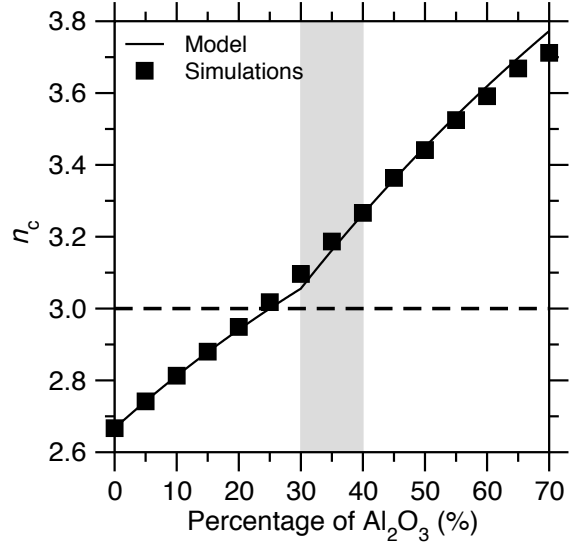

**FIG. 4:** Topological constraint model for $(Na_2O)_{30}(SiO_2)_{70-x}(Al_2O_3)_x$. The solid line is constructed based on the number of constraints calculated using theoretical atomic fractions, and the squares are constructed based on the number of constraints ($n_c$) calculated using simulated atomic fractions. The dotted line indicates where the $n_c$ is exactly balanced by the number of degrees of freedom. The grey area indicates the region where glass is isostatic.

The number of constraints per atom ($n_c$) as a function of percentage $Al_2O_3$ is shown in FIG. 4. Simulation data points are obtained by multiplying the simulated fractions of each atomic sub-species with their respective number of constraints per atom. The close agreement between the number of topological constraints predicted by theoretical atomic fractions and the number of topological constraints calculated by simulated atomic fractions suggest that the simulated fractions of each type of oxygen (NBO, BO and TO) and aluminum atoms ($Al^{IV}$ and $Al^V$) closely reflect the theoretical fractions. The $n_c$ is observed to increase as the percentage of alumina increases, and $n_c$ increases more steeply after 30% alumina due to the formation of TO. At 25% alumina, the $n_c$ is exactly balanced by the number of degrees of freedom, i.e. the glass is isostatic. The proposed topological constraint model can be indirectly verified by experiments since isostatic glasses were found to have the best glass forming abilities [8]. Indeed, alumina rich glasses are extremely difficult to form in the laboratory [62], which is why molecular dynamics simulations are relied upon to study per-aluminous aluminosilicates.

## D. Signatures of flexibility and stress

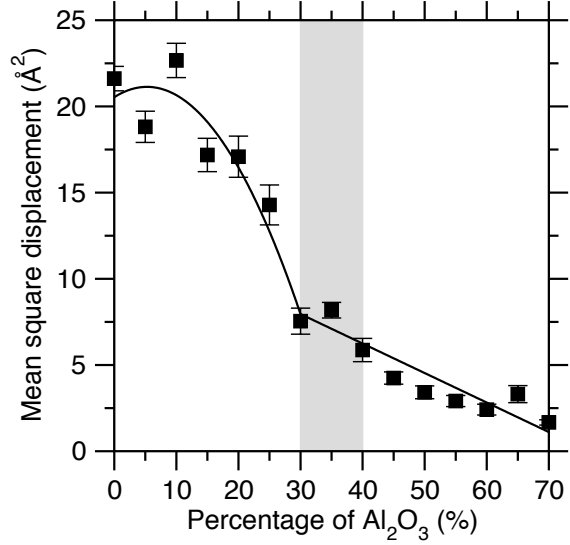

**FIG. 5**: Mean square displacement of all atoms after an energy bump of 194 meV and a duration of 2 ns, as a function of percentage $Al_2O_3$. The line serves as a guide for the eye. The gray area indicates the region of rigidity transition.

The mean square displacement (MSD) of all atoms, a signature of network flexibility, after an energy bump of 194 meV and a duration of 2 ns is shown in FIG. 5. The effect of the energy bump is to provide the atoms with enough kinetic energy to move and displace permanently, but not to the extent of melting. This method has been used in previous studies [43,63]. Mean square displacements of up to 23 $Å^2$ are observed, which imply that atoms are permanently displaced. This occurs when atoms are provided with enough energy to jump over energy barriers, moving from one minimum in the enthalpy landscape to other minimums. MSD is observed to decrease slightly from 0% to 25% alumina, but a sudden drop in MSD is observed at 30% alumina. High MSD values prior to 30% alumina suggest that those glass compositions are flexible and have internal degrees of freedom. The sharp drop at 30% signifies a rigidity transition at the percolation threshold [64], at which long-range connectivity in the glassy network is achieved. Hence, this paper defines 30% alumina as the lower bound of rigidity transition. MSD is observed to decrease linearly from 30% to 70% alumina as the number of constraints continues to increase due to the addition of alumina.

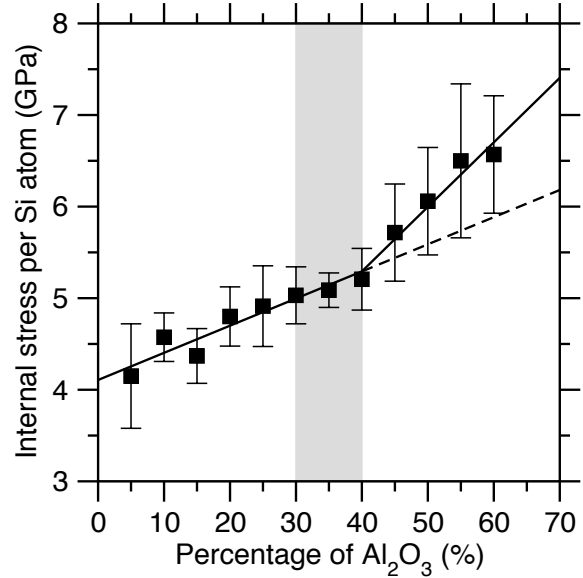

**FIG. 6:** Internal stress per silicon atom as a function of percentage $Al_2O_3$. The lines serve as a guide for the eye. The gray area indicates the region of rigidity transition.

For a pure sodium silicate glass ($x = 0\%$), although it is in the flexible regime, there exists some internal stress due to residual thermal stress from quenching [41,65]. In FIG. 6, the internal stress per silicon atom is observed to increase as the percentage of alumina increases, which corresponds to an increase in $n_c$. Moreover, an increase in slope is observed at 40% alumina, which indicates the presence of additional internal stress due to the network becoming over-constrained. Hence, this paper defines 40% as the point of transition into the stressed-rigid domain.

## IV. CONCLUSIONS

Altogether, this study provides a topological model for sodium aluminosilicate glasses, which can be used to predict the composition wherein the glass is expected to exhibit a flexible-to-stressed–rigid transition. We find that the enumeration of the topological constraints is independently supported by tracking the onset of flexibility and internal stress within the atomic network. As a major outcome of this study, we find that, as the atomic connectivity increases, the glass first exhibits a flexible-to-rigid transition (i.e., a loss of atomic mobilities) and then an unstressed-to-stressed transition (i.e., an onset of internal stress). The fact that these two transitions do not occur at the same threshold composition effectively defines a range of compositions wherein the glass is isostatic, that is, rigid but free of internal stress [66–69].

**REFERENCES**
[1] D.K. Hale, Strengthening of Silicate Glasses by Ion Exchange, Nature. 217 (1968) 1115–1118. doi:10.1038/2171115a0.
[2] A.R. Cooper, D.A. Krohn, Strengthening of Glass Fibers: 11, Ion Exchange, Journal of the American Ceramic Society. 52 (1969) 665–669. doi:10.1111/j.1151-2916.1969.tb16073.x.

[3] A.K. Varshneya, Chemical Strengthening of Glass: Lessons Learned and Yet To Be Learned, International Journal of Applied Glass Science. 1 (2010) 131–142. doi:10.1111/j.2041-1294.2010.00010.x.

[4] A.K. Varshneya, The Physics of Chemical Strengthening of Glass: Room for a New View, Journal of Non-Crystalline Solids. 356 (2010) 2289–2294. doi:10.1016/j.jnoncrysol.2010.05.010.

[5] R.C. Welch, J.R. Smith, M. Potuzak, X. Guo, B.F. Bowden, T.J. Kiczenski, D.C. Allan, E.A. King, A.J. Ellison, J.C. Mauro, Dynamics of Glass Relaxation at Room Temperature, Physical Review Letters. 110 (2013). doi:10.1103/PhysRevLett.110.265901.

[6] A. Ellison, I.A. Cornejo, Glass Substrates for Liquid Crystal Displays, International Journal of Applied Glass Science. 1 (2010) 87–103. doi:10.1111/j.2041-1294.2010.00009.x.

[7] D. Käfer, M. He, J. Li, M.S. Pambianchi, J. Feng, J.C. Mauro, Z. Bao, Ultra-Smooth and Ultra-Strong Ion-Exchanged Glass as Substrates for Organic Electronics, Advanced Functional Materials. 23 (2013) 3233–3238. doi:10.1002/adfm.201202009.

[8] J.C. Phillips, Topology of Covalent Non-Crystalline Solids I: Short-Range Order in Chalcogenide Alloys, Journal of Non-Crystalline Solids. 34 (1979) 153–181. doi:10.1016/0022-3093(79)90033-4.

[9] M.F. Thorpe, Continuous Deformations in Random Networks, Journal of Non-Crystalline Solids. 57 (1983) 355–370. doi:10.1016/0022-3093(83)90424-6.

[10] J.C. Mauro, Topological Constraint Theory of Glass, American Ceramic Society Bulletin. 90 (2011) 31.

[11] M. Bauchy, Topological Constraints and Rigidity of Network Glasses from Molecular Dynamics Simulations, Am. Ceram. Soc. Bull. 91 (2012) 34.

[12] M. Micoulaut, Y. Yue, Material Functionalities from Molecular Rigidity: Maxwell's Modern Legacy, MRS Bulletin. 42 (2017) 18–22. doi:10.1557/mrs.2016.298.

[13] M. Bauchy, Nanoengineering of concrete via topological constraint theory, MRS Bulletin. 42 (2017) 50–54. doi:10.1557/mrs.2016.295.

[14] H. Liu, T. Du, N.M.A. Krishnan, H. Li, M. Bauchy, Topological optimization of cementitious binders: Advances and challenges, Cement and Concrete Composites. (2018). doi:10.1016/j.cemconcomp.2018.08.002.

[15] M. Bauchy, Deciphering the atomic genome of glasses by topological constraint theory and molecular dynamics: A review, Computational Materials Science. 159 (2019) 95–102. doi:10.1016/j.commatsci.2018.12.004.

[16] M.V. Chubynsky, M.-A. Brière, N. Mousseau, Self-Organization with Equilibration: A Model for the Intermediate Phase in Rigidity Percolation, Phys. Rev. E. 74 (2006) 016116. doi:10.1103/PhysRevE.74.016116.

[17] D.M. Zirl, S.H. Garofalini, Structure of Sodium Aluminosilicate Glasses, Journal of the American Ceramic Society. 73 (1990) 2848–2856. doi:10.1111/j.1151-2916.1990.tb06685.x.

[18] Y. Xiang, J. Du, M.M. Smedskjaer, J.C. Mauro, Structure and Properties of Sodium Aluminosilicate Glasses from Molecular Dynamics Simulations, The Journal of Chemical Physics. 139 (2013) 044507. doi:10.1063/1.4816378.

[19] M. Ren, J.Y. Cheng, S.P. Jaccani, S. Kapoor, R.E. Youngman, L. Huang, J. Du, A. Goel, Composition – structure – property relationships in alkali aluminosilicate glasses: A combined experimental – computational approach towards designing functional glasses, Journal of Non-Crystalline Solids. 505 (2019) 144–153. doi:10.1016/j.jnoncrysol.2018.10.053.

[20] J.P. Hamilton, C.G. Pantano, Effects of Glass Structure on the Corrosion Behavior of Sodium-Aluminosilicate Glasses, Journal of Non-Crystalline Solids. 222 (1997) 167–174. doi:10.1016/S0022-3093(97)90110-1.

[21] D.A. McKeown, F.L. Galeener, G.E. Brown, Raman Studies of Al Coordination in Silica-Rich Sodium Aluminosilicate Glasses and Some Related Minerals, Journal of Non-Crystalline Solids. 68 (1984) 361–378. doi:10.1016/0022-3093(84)90017-6.

[22] J. Du, Molecular Dynamics Simulations of the Structure and Properties of Low Silica Yttrium Aluminosilicate Glasses, Journal of the American Ceramic Society. 92 (2009) 87–95. doi:10.1111/j.1551-2916.2008.02853.x.

[23] M. Wang, M.M. Smedskjaer, J.C. Mauro, G. Sant, M. Bauchy, Topological Origin of the Network Dilation Anomaly in Ion-Exchanged Glasses, Phys. Rev. Applied. 8 (2017) 054040. doi:10.1103/PhysRevApplied.8.054040.

[24] C. Hermansen, X. Guo, R.E. Youngman, J.C. Mauro, M.M. Smedskjaer, Y. Yue, Structure-Topology-Property Correlations of Sodium Phosphosilicate Glasses, J. Chem. Phys. 143 (2015) 064510. doi:10.1063/1.4928330.

[25] M.M. Smedskjaer, J.C. Mauro, Y. Yue, Prediction of Glass Hardness Using Temperature-Dependent Constraint Theory, Phys. Rev. Lett. 105 (2010) 115503. doi:10.1103/PhysRevLett.105.115503.

[26] S. Plimpton, Fast Parallel Algorithms for Short-Range Molecular Dynamics, Journal of Computational Physics. 117 (1995) 1–19. doi:10.1006/jcph.1995.1039.

[27] J. Du, Challenges in Molecular Dynamics Simulations of Multicomponent Oxide Glasses, in: C. Massobrio, C. Du, M. Bernasconi, P.S. Salmon (Eds.), Molecular Dynamics Simulations of Disordered Materials, Springer International Publishing, 2015: pp. 157–180.

[28] J. Du, A.N. Cormack, The Medium Range Structure of Sodium Silicate Glasses: A Molecular Dynamics Simulation, Journal of Non-Crystalline Solids. 349 (2004) 66–79. doi:10.1016/j.jnoncrysol.2004.08.264.

[29] X. Li, W. Song, K. Yang, N.M.A. Krishnan, B. Wang, M.M. Smedskjaer, J.C. Mauro, G. Sant, M. Balonis, M. Bauchy, Cooling Rate Effects in Sodium Silicate Glasses: Bridging the Gap Between Molecular Dynamics Simulations and Experiments, J. Chem. Phys. 147 (2017) 074501. doi:10.1063/1.4998611.

[30] M. Bauchy, Structural, vibrational, and thermal properties of densified silicates: Insights from molecular dynamics, J. Chem. Phys. 137 (2012) 044510. doi:10.1063/1.4738501.

[31] M. Bauchy, M. Micoulaut, From Pockets to Channels: Density-Controlled Diffusion in Sodium Silicates, Physical Review B. 83 (2011) 184118. doi:10.1103/PhysRevB.83.184118.

[32] Z. Liu, Y. Hu, X. Li, W. Song, S. Goyal, M. Micoulaut, M. Bauchy, Glass relaxation and hysteresis of the glass transition by molecular dynamics simulations, Phys. Rev. B. 98 (2018) 104205. doi:10.1103/PhysRevB.98.104205.

[33] S. Nosé, A Molecular Dynamics Method for Simulations in the Canonical Ensemble, Molecular Physics. 52 (1984) 255–268. doi:10.1080/00268978400101201.

[34] W.G. Hoover, Canonical Dynamics: Equilibrium Phase-Space Distributions, Phys. Rev. A. 31 (1985) 1695–1697. doi:10.1103/PhysRevA.31.1695.

[35] MATLAB, The MathWorks, Inc., Natick, Massachusetts, United States, 2017.

[36]  A. Stukowski, Visualization and Analysis of Atomistic Simulation Data with OVITO–the Open Visualization Tool, Modelling Simul. Mater. Sci. Eng. 18 (2010) 015012. doi:10.1088/0965-0393/18/1/015012.

[37]  M. Wang, N.M. Anoop Krishnan, B. Wang, M.M. Smedskjaer, J.C. Mauro, M. Bauchy, A new transferable interatomic potential for molecular dynamics simulations of borosilicate glasses, Journal of Non-Crystalline Solids. 498 (2018) 294–304. doi:10.1016/j.jnoncrysol.2018.04.063.

[38]  K. Yang, A. Kachmar, B. Wang, N.M.A. Krishnan, M. Balonis, G. Sant, M. Bauchy, New insights into the atomic structure of amorphous TiO2 using tight-binding molecular dynamics, J. Chem. Phys. 149 (2018) 094501. doi:10.1063/1.5042783.

[39]  L.M. Adams, J.L. Nazareth, in: Linear and Nonlinear Conjugate Gradient-Related Methods, SIAM, 1996.

[40]  A.P. Thompson, S.J. Plimpton, W. Mattson, General Formulation of Pressure and Stress Tensor for Arbitrary Many-Body Interaction Potentials Under Periodic Boundary Conditions, J Chem Phys. 131 (2009) 154107. doi:10.1063/1.3245303.

[41]  X. Li, Quantifying the Internal Stress in Over-Constrained Glasses by Molecular Dynamics Simulations, (n.d.).

[42]  Y. Yu, M. Wang, M.M. Smedskjaer, J.C. Mauro, G. Sant, M. Bauchy, Thermometer Effect: Origin of the Mixed Alkali Effect in Glass Relaxation, Phys. Rev. Lett. 119 (2017) 095501. doi:10.1103/PhysRevLett.119.095501.

[43]  B. Wang, N.M.A. Krishnan, Y. Yu, M. Wang, Y. Le Pape, G. Sant, M. Bauchy, Irradiation-Induced Topological Transition in Sio2: Structural Signature of Networks' Rigidity, Journal of Non-Crystalline Solids. 463 (2017) 25–30. doi:10.1016/j.jnoncrysol.2017.02.017.

[44]  H. Liu, S. Dong, L. Tang, N.M.A. Krishnan, G. Sant, M. Bauchy, Effects of Polydispersity and Disorder on the Mechanical Properties of Hydrated Silicate Gels, Journal of the Mechanics and Physics of Solids. 122 (2019) 555–565. doi:10.1016/j.jmps.2018.10.003.

[45]  Y. Yu, M. Wang, N.M. Anoop Krishnan, M.M. Smedskjaer, K. Deenamma Vargheese, J.C. Mauro, M. Balonis, M. Bauchy, Hardness of silicate glasses: Atomic-scale origin of the mixed modifier effect, Journal of Non-Crystalline Solids. 489 (2018) 16–21. doi:10.1016/j.jnoncrysol.2018.03.015.

[46]  Y. Yu, J.C. Mauro, M. Bauchy, Stretched exponential relaxation of glasses: Origin of the mixed-alkali effect, American Ceramic Society Bulletin. 96 (2017) 34–36.

[47]  Y. Yu, B. Wang, M. Wang, G. Sant, M. Bauchy, Reactive Molecular Dynamics Simulations of Sodium Silicate Glasses — Toward an Improved Understanding of the Structure, Int J Appl Glass Sci. 8 (2017) 276–284. doi:10.1111/ijag.12248.

[48]  M. Bauchy, Structural, vibrational, and elastic properties of a calcium aluminosilicate glass from molecular dynamics simulations: The role of the potential, The Journal of Chemical Physics. 141 (2014) 024507. doi:10.1063/1.4886421.

[49]  A.K. Varshneya, Fundamentals of Inorganic Glasses, Elsevier, 1994.

[50]  J.F. Stebbins, Z. Xu, NMR Evidence for Excess Non-Bridging Oxygen in an Aluminosilicate Glass, Nature. 390 (1997) 60–62. doi:10.1038/36312.

[51]  N. Jakse, M. Bouhadja, J. Kozaily, J.W.E. Drewitt, L. Hennet, D.R. Neuville, H.E. Fischer, V. Cristiglio, A. Pasturel, Interplay Between Non-Bridging Oxygen, Triclusters, and Fivefold Al Coordination in Low Silica Content Calcium Aluminosilicate Melts, Applied Physics Letters. 101 (2012) 201903. doi:10.1063/1.4766920.

[52] M. Bauchy, M. Micoulaut, M. Celino, S. Le Roux, M. Boero, C. Massobrio, Angular rigidity in tetrahedral network glasses with changing composition, Phys. Rev. B. 84 (2011) 054201. doi:10.1103/PhysRevB.84.054201.

[53] M. Wang, B. Wang, T.K. Bechgaard, J.C. Mauro, S.J. Rzoska, M. Bockowski, M.M. Smedskjaer, M. Bauchy, Crucial effect of angular flexibility on the fracture toughness and nano-ductility of aluminosilicate glasses, Journal of Non-Crystalline Solids. 454 (2016) 46–51. doi:10.1016/j.jnoncrysol.2016.10.020.

[54] M. Bauchy, M. Micoulaut, Atomic scale foundation of temperature-dependent bonding constraints in network glasses and liquids, Journal of Non-Crystalline Solids. 357 (2011) 2530–2537. doi:10.1016/j.jnoncrysol.2011.03.017.

[55] M. Micoulaut, M. Bauchy, H. Flores-Ruiz, Topological Constraints, Rigidity Transitions, and Anomalies in Molecular Networks, Molecular Dynamics Simulations of Disordered Materials. (2015) 275–311.

[56] G. Gutiérrez, B. Johansson, Molecular Dynamics Study of Structural Properties of Amorphous Al2o3, Phys. Rev. B. 65 (2002) 104202. doi:10.1103/PhysRevB.65.104202.

[57] E. Dupree, R.F. Pettifer, Determination of the Si–O–Si Bond Angle Distribution in Vitreous Silica by Magic Angle Spinning NMR, Nature. 308 (1984) 523–525. doi:10.1038/308523a0.

[58] F. Angeli, J.-M. Delaye, T. Charpentier, J.-C. Petit, D. Ghaleb, P. Faucon, Investigation of Al–O–Si Bond Angle in Glass by 27Al 3q-MAS NMR and Molecular Dynamics, Chemical Physics Letters. 320 (2000) 681–687. doi:10.1016/S0009-2614(00)00277-3.

[59] R. Oestrike, W. Yang, R.J. Kirkpatrick, R.L. Hervig, A. Navrotsky, B. Montez, High-Resolution 23Na, 27Al and 29Si NMR Spectroscopy of Framework Aluminosilicate Glasses, Geochimica et Cosmochimica Acta. 51 (1987) 2199–2209. doi:10.1016/0016-7037(87)90269-9.

[60] Y. Yu, B. Wang, M. Wang, G. Sant, M. Bauchy, Revisiting silica with ReaxFF: Towards improved predictions of glass structure and properties via reactive molecular dynamics, Journal of Non-Crystalline Solids. 443 (2016) 148–154. doi:10.1016/j.jnoncrysol.2016.03.026.

[61] M. Bauchy, M.J. Abdolhosseini Qomi, C. Bichara, F.-J. Ulm, R.J.-M. Pellenq, Nanoscale Structure of Cement: Viewpoint of Rigidity Theory, J. Phys. Chem. C. 118 (2014) 12485–12493. doi:10.1021/jp502550z.

[62] A. Rosenflanz, M. Frey, B. Endres, T. Anderson, E. Richards, C. Schardt, Bulk Glasses and Ultrahard Nanoceramics Based on Alumina and Rare-Earth Oxides, Nature. 430 (2004) 761–764. doi:10.1038/nature02729.

[63] N.M.A. Krishnan, B. Wang, Y. Yu, Y. Le Pape, G. Sant, M. Bauchy, Enthalpy Landscape Dictates the Irradiation-Induced Disordering of Quartz, Phys. Rev. X. 7 (2017) 031019. doi:10.1103/PhysRevX.7.031019.

[64] J.C. Phillips, M.F. Thorpe, Constraint Theory, Vector Percolation and Glass Formation, Solid State Communications. 53 (1985) 699–702. doi:10.1016/0038-1098(85)90381-3.

[65] X. Yang, R.J. Young, Model Ceramic Fibre-Reinforced Glass Composites: Residual Thermal Stresses, Composites. 25 (1994) 488–493. doi:10.1016/0010-4361(94)90174-0.

[66] P. Boolchand, D.G. Georgiev, B. Goodman, Discovery of the intermediate phase in chalcogenide glasses, Journal of Optoelectronics and Advanced Materials. 3 (2001) 703–720.

[67]     P. Boolchand, M. Bauchy, M. Micoulaut, C. Yildirim, Topological Phases of Chalcogenide Glasses Encoded in the Melt Dynamics, Physica Status Solidi (B). 255 (2018) 1800027. doi:10.1002/pssb.201800027.

[68]     N.M.A. Krishnan, B. Wang, G. Sant, J.C. Phillips, M. Bauchy, Revealing the Effect of Irradiation on Cement Hydrates: Evidence of a Topological Self-Organization, ACS Appl. Mater. Interfaces. 9 (2017) 32377–32385. doi:10.1021/acsami.7b09405.

[69]     M. Bauchy, M. Wang, Y. Yu, B. Wang, N.M.A. Krishnan, E. Masoero, F.-J. Ulm, R. Pellenq, Topological Control on the Structural Relaxation of Atomic Networks under Stress, Phys. Rev. Lett. 119 (2017) 035502. doi:10.1103/PhysRevLett.119.035502.